\documentclass[prb,aps,twocolumn,amsmath,amssymb,showpacs,superscriptaddress]{revtex4}
\usepackage{graphicx}
\usepackage{psfrag}
\usepackage{color}
\usepackage{subfigure}
\usepackage{amsmath}
\definecolor{dred}{rgb}{0.7,0.0,0.0}
\allowdisplaybreaks 
 
\begin{document}

%
%

\title{Constraints Imposed by Symmetry on Pairing Operators for the Pnictides}

\author{Xiaoyu Wang}
 
\affiliation{Department of Physics and Astronomy, The University of
  Tennessee, Knoxville, TN 37996} 
\affiliation{Materials Science and Technology Division, Oak Ridge
  National Laboratory, Oak Ridge, TN 32831} 
\affiliation{Department of Electronic Communication and Engineering, 
Shanghai Jiao Tong University, Shanghai 200240}

\author{Maria Daghofer}

\affiliation{IFW Dresden, P.O. Box 27 01 16, D-01171 Dresden, Germany}
\affiliation{Department of Physics and Astronomy, The University of
  Tennessee, Knoxville, TN 37996} 
\affiliation{Materials Science and Technology Division, Oak Ridge
  National Laboratory, Oak Ridge, TN 32831} 

\author{Andrew Nicholson}
\author{Adriana Moreo}
\author{Michael Guidry}
\author{Elbio Dagotto}
 
\affiliation{Department of Physics and Astronomy, The University of
  Tennessee, Knoxville, TN 37996} 
\affiliation{Materials Science and Technology Division, Oak Ridge
  National Laboratory, Oak Ridge, TN 32831}

\date{\today}

\begin{abstract}

Considering model Hamiltonians 
that respect the symmetry properties of the pnictides, 
it is argued
that pairing interactions that couple electrons at different orbitals 
with an orbital-dependent pairing strength inevitably lead to interband pairing 
matrix elements, at least in some regions of the Brillouin zone. Such
interband pairing has not been considered of relevance in multiorbital
systems in previous investigations.
It is also observed that if, instead, 
a purely intraband pairing interaction is postulated, this requires 
that the pairing operator has the form $\Delta^{\dagger}({\bf k})=f({\bf k})
\sum_{\alpha} d^{\dagger}_{{\bf k},\alpha,\uparrow}d^{\dagger}_{-{\bf k},
\alpha,\downarrow}$ where $\alpha$ labels the orbitals considered in the model
and $f({\bf k})$ arises from the spatial location of the coupled electrons or 
holes. This means that the gaps at two different Fermi surfaces involving momenta
${\bf k}_F$ and ${\bf k}'_F$ can only differ by the ratio $f({\bf k}_F)/
f({\bf k}'_F)$ and that electrons in different orbitals must be subject to the 
same pairing attraction, thus requiring fine tuning.  
These results suggest that previously neglected 
interband pairing tendencies could actually be of relevance in 
a microscopic description of the pairing mechanism in the pnictides.


\end{abstract}
 
\pacs{75.50.Pp, 71.10.Fd, 72.25.Dc}
 
\maketitle

\section {Introduction} 

The discovery of superconductivity in the iron-based 
pnictides\cite{Fe-SC,chen1,chen2,wen,chen3,ren1,55,ren2} has opened a new 
active direction of research in the quest to understand high critical
temperature (high-$T_c$) superconductors. 
Experiments are showing
that the pnictides share several properties with the high-$T_c$ 
cuprates, such as the order of magnitude of the critical 
temperature,\cite{Fe-SC} the existence of magnetic order in some of the parent 
compounds,\cite{sdw,neutrons1,neutrons2,neutrons3,neutrons4} and a possible
exotic pairing mechanism.\cite{phonon0}
However, there are differences in several aspects as well: the parent compound 
is a (bad) metal instead of a Mott
insulator,\cite{sdw,neutrons1,neutrons2,neutrons3,neutrons4} and several 
orbitals, as opposed to only one, 
have to be considered in order to reproduce the Fermi surface, which consists 
of hole and electron pockets.\cite{first,xu,cao,fang2}
In addition, 
while clear experimental evidence and theoretical calculations indicate that 
the pairing state in the cuprates is nodal and has $d$-wave 
symmetry,\cite{review} the 
properties of the pairing operator in the pnictides have not yet been 
established. Experimentally, several angle resolved photoemission 
(ARPES) studies\cite{arpes,arpes2,arpes3,arpes22,hsieh,rossle} show 
constant nodeless gaps on all Fermi surfaces (FSs), but 
evidence for the existence of nodal 
gaps has been reported in many transport measurements as well.\cite{nodal1,nodal2,nodal3,Ahilan,nakai,Grafe,Y.Wang,matano,mukuda,millo,wang-nodes} 
It has been argued that the 
properties of the gap may be material dependent or that a nodal gap may be 
rendered nodeless by disorder,\cite{disorder} but a consensus has not been
reached.

The goal of this paper is to understand the constraints that symmetries and the number of 
active degrees of freedom in  
the pnictides impose on the possible pairing operators. We will consider a {\it five}
orbital model that retains the five $d$ orbitals of each of the two Fe atoms in the unit 
cell of the FeAs planes.\cite{graser} Employing mean-field approximations
we will discuss comparisons with results
obtained in models with three\cite{our3} and 
two\cite{Daghofer:2009p1970,moreo,rong} 
orbitals that can be studied numerically.

The organization of the paper is as follows: in Section~\ref{model} we present
the five-orbital Hamiltonian written in terms of SU(5) 
$5 \times 5$ matrices to properly identify
its symmetries. The pairing operators
are explicitly discussed in Section~\ref{pairingo}. Section~\ref{disc} is devoted to a 
discussion of the results, while conclusions are provided in 
Section~\ref{concl}.

\section{Model}\label{model}

To construct the possible pairing operators allowed by the lattice 
and orbital symmetries, we will follow the approach already described
in detail for the case of a three-orbital model.\cite{our3}
The first step involves the rewriting of the tight-binding Hamiltonian
for the five-orbital model, for instance as  
provided in Ref.~\onlinecite{graser}, 
in terms of some of the 25 $5\times5$ matrix generators of $U(5)\supset
U(1) \times SU(5)$, which consist of the unit $5\times5$ matrix and 24
$5\times 5$ matrix generators of $SU(4)$. Since most of the elements
of these matrices are zero, instead of providing for the reader 
this large number of matrices explicitly,
it is more convenient to simply describe them verbally.
For $i=1$ up to 8, the $\lambda_i$ matrices are 
given by the well-known eight Gell-Mann matrices\cite{our3} in the upper left-hand corner 
while the rest of the elements are equal to zero. The matrices with $i=9$ $(1,4)$, 
11 $(1,5)$, 13 $(2,4)$, 15 $(2,5)$, 17 $(3,4)$, 19 $(3,5)$, and 21 $(4,5)$ are symmetric 
and have only two elements equal to 1 while the rest of the elements are 0, with $(i,j)$ indicating
the position of one of the non-zero elements (by symmetry the other element can be identified).
 The matrices with $i=10$ $(1,4)$, 
12 $(1,5)$, 14 $(2,4)$, 16 $(2,5)$, 18 $(3,4)$, 20 $(3,5)$, and 22 $(4,5)$ are Hermitian 
and have two elements equal to $\pm i$ and the rest are 0, with $(i,j)$ 
indicating the position of the non-zero element equal to $-i$. Finally, 
$\lambda_{23}$ is diagonal with nonzero elements (4,4) and (5,5) equal to 1
and -1, and $\lambda_{24}$ is diagonal with nonzero elements 
$(1,1)=(2,2)=(3,3)=\frac{2}{\sqrt{15}}$, $(4,4)=\frac{-3}{\sqrt{15}}$ and 
$(5,5)=\frac{-1}{\sqrt{15}}$.

The Hamiltonian then becomes
\begin{equation}
H_{\rm TB}( {\bf k})=\sum_{ {\bf k},\sigma}\Phi^{\dagger}_{ {\bf k},\sigma}
\zeta_{ {\bf k}}\Phi_{ {\bf k},\sigma},
\label{1}
\end{equation}
where $\Phi^{\dagger}_{ {\bf k},\sigma}=
(d^{\dagger}_{1}({\bf k}),d^{\dagger}_{2}({\bf k}),
d^{\dagger}_{3}({\bf k}),d^{\dagger}_{4}({\bf k}),
d^{\dagger}_{5}({\bf k}))_{\sigma}$ and
\begin{equation}
\zeta_{ {\bf k}}=\sum_{i=0}^{24} a_i\lambda_i,
\label{2}
\end{equation}
where the coefficients $a_i$ are presented in Table~\ref{table1}.
The functions $\xi_{ij}$ appearing in this Table are provided in the Appendix.
The on-site energies for each orbital are given by
$\epsilon_1=\epsilon_2=0.13$ eV, $\epsilon_3=-0.22$ eV, $\epsilon_4= 0.3$ eV, 
and
$\epsilon_5= -0.211$ eV.\cite{graser} In addition, $e_i=\xi_{ii}+\epsilon_i$, 
and the chemical potential is 0. The
index-to-orbital correspondence is the following: (1) $xz$, (2) $yz$, (3) $x^2-y^2$,
(4) $xy$, and (5) $3z^2-r^2$.\cite{graser}

 \begin{table}
\caption{Coefficients for the $\lambda_i$ matrices in Eq.~\ref{2} and the 
irreducible representation of $D_{\rm 4h}$ according to which they transform.
The $\xi_{ij}$ are provided in the Appendix.}
 \begin{tabular}{|c|c|c|}\hline
$i$ & $a_i$ & IR \\ 
\hline
0 &$\frac{\sum_i e_i}{5}$ & $A_{\rm 1g}$ \\
1 & $\xi_{12}$&$B_{\rm 2g}$\\
2 &$0$&\\
3 &$\frac{e_1-e_2}{2}$&$B_{\rm 1g}$\\
4 &$0$&\\
5 &$i\xi_{13}$&$E_{\rm g}$\\
6 &$0$&\\
7 &$i\xi_{23}$&$E_{\rm g}$\\
8 &$\frac{e_1+e_2-2e_3}{2\sqrt{3}}$&$A_{\rm 1g}$\\
9 &$0$&\\
10 &$i\xi_{14}$&$E_{\rm g}$\\
11 &$0$&\\
12 &$i\xi_{15}$&$E_{\rm g}$\\
13 &$0$&\\
14 &$i\xi_{24}$&$E_{\rm g}$\\
15 &$0$&\\
16 &$i\xi_{25}$&$E_{\rm g}$\\
17 &$\xi_{34}$&$A_{\rm 2g}$\\
18 &$0$&\\
19 &$\xi_{35}$&$B_{\rm 1g}$\\
20 &$0$&\\
21 &$\xi_{45}$&$B_{\rm 2g}$\\
22 &$0$&\\
23 &$\frac{e_4-e_5}{2}$&$A_{\rm 1g}$\\
24 &$\frac{\sqrt{15}}{5}\left(\frac{e_1 +e_2 +e_3}{3}-\frac{e_4 +e_5}{2}\right)$&$A_{\rm 1g}$\\
\hline
\end{tabular}
\label{table1}
 \end{table}
The symmetry operations that leave invariant the 
Fe-As planes can be mapped on the elements of the point 
group $D_{\rm 4h}$\cite{eschrig,zhou} 
and, 
thus, the Hamiltonian has to remain invariant under all the operations of this 
group, which means that it must transform according to the irreducible 
representation $A_{\rm 1g}$. This allows us to assign an
irreducible representation to each of the $\lambda_i$ matrices 
and, thus, classify possible pairing operators according to 
their symmetry.\cite{wan,our3} It is important to realize that due to the strong hybridization
among the orbitals, evident in all the terms in Eq.~\ref{1} where $\xi_{ij}$ with $i\ne j$ appear, a 
proper characterization of the pairing operators by symmetry cannot be accomplished using only the band representation. 


\section{Pairing Operators} \label{pairingo}

In order to construct the allowed 
pairing operators in multiorbital models the symmetries of the 
spatial, spin, and orbital contributions needs to be considered. Ignoring the 
orbital symmetry may lead to problems similar to those encountered in the 
early days of the study of magnetism when the spin contribution 
to the electronic wave functions was 
not included. Note that from the point of view of the orbital 
symmetry some of the previous 
efforts on superconductivity in multiorbital systems, such as in 
Ref.~\onlinecite{suhl}, are effectively dealing with non-hybridized 
``$s$-orbitals''. Thus, those results cannot be straightforwardly applied to a 
system with strongly hybridized non-$s$ orbitals. In fact, it was observed that
the general form of a spin-singlet pairing operator in this case is 
given by\cite{ib}
\begin{equation}
\Delta^{\dagger}({\bf k})= f_i({\bf k})
(\lambda_i)_{\alpha,\beta}
(d^{\dagger}_{{\bf k},\alpha,\uparrow} d^{\dagger}_{{\bf -k},\beta,\downarrow}
-d^{\dagger}_{{\bf k},\beta,\uparrow} 
d^{\dagger}_{{\bf -k},\alpha,\downarrow}),
\label{3}
\end{equation}
\noindent where a sum over repeated indices $\alpha$ and $\beta$ 
is implied; the operators 
$d$ have been defined above, and $f({\bf k})$ is the form factor
that transforms according to one of the irreducible representations of 
the crystal's symmetry group.\cite{our3} The form factors that will be 
considered in this work and their corresponding 
irreducible representations of $D_{\rm 4h}$ 
(according to which they transform) are presented in Table~\ref{table2}. The 
index $i$ in Eq.~(\ref{3})indicates that different form factors may be needed if matrices 
$\lambda_i$ with different symmetries are combined.

\begin{table}
\caption{Form factors $f({\bf k})$ for pairs up to distance (1,1) classified
according to their symmetry under $D_{\rm 4h}$ operations.}
 \begin{tabular}{|l|c|r|} \hline
 \# & $f({\bf k})$ & IR \\
\hline
1 & 1 & $A_{\rm 1g}$\\
2 & $\cos(k_x)+\cos(k_y)$ & $A_{\rm 1g}$\\
3 & $\cos(k_x)\cos(k_y)$ & $A_{\rm 1g}$ \\
4 & $\cos(k_x)-\cos(k_y)$ & $B_{\rm 1g}$\\
5 & $\sin(k_x)\sin(k_y)$ & $B_{\rm 2g}$ \\
6 & ($\sin(k_x),\sin(k_y)$) & $E_{\rm g}$\\
7 & $(\sin(k_x)\cos(k_y),\sin(k_y)\cos(k_x))$& $E_{\rm g}$\\
\hline
\end{tabular}
\label{table2}
\end{table}

\begin{figure}[thbp]
\begin{center}
\includegraphics[width=8cm,trim = 30 30 25 35,clip,angle=0]{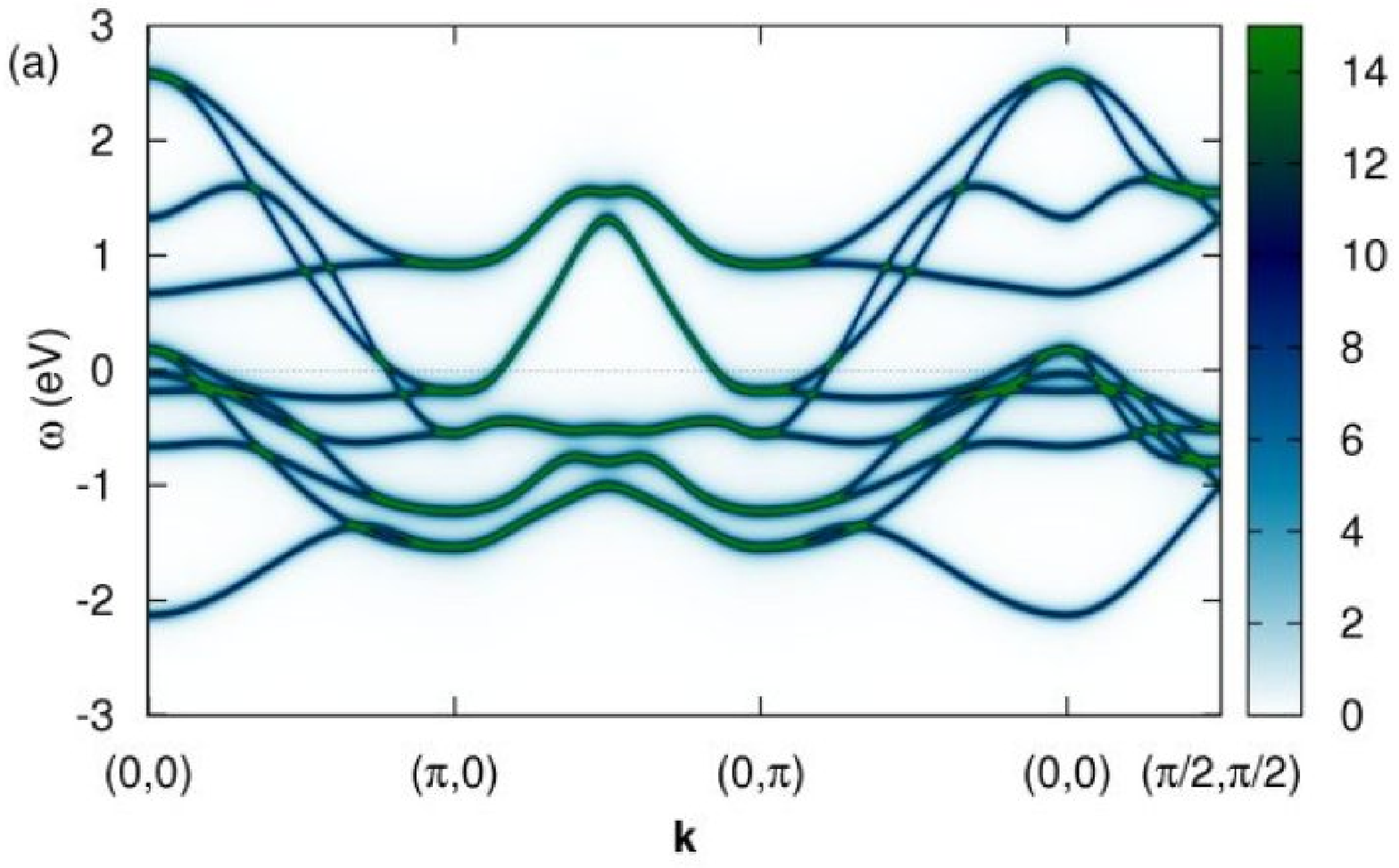}
\includegraphics[width=8cm,trim = 30 30 25 35,clip,angle=0]{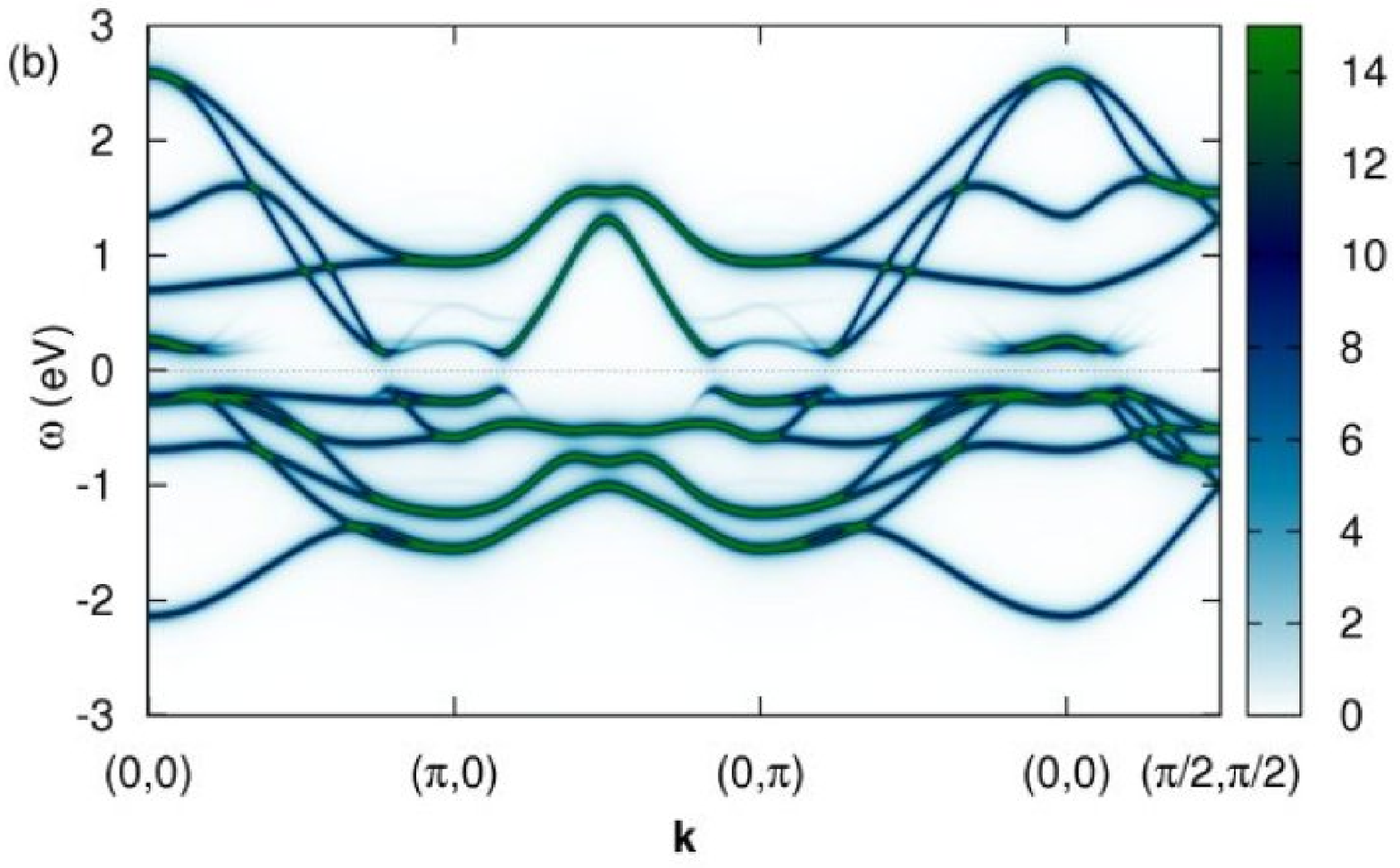}
\vskip 0.3cm
\caption{(Color online) The intensity of the points represents the values
of the
spectral function $A({\bf{k},\omega})$ for the five-orbital model with pairing
interaction (a) $V=0$; (b) $V=0.2$, and for the $S \pm$ pairing operator given
in the text. The results are shown in the reduced BZ.}
\label{akwv0spm}
\end{center}
\end{figure}

As in the case of the two\cite{wan} 
and three\cite{our3} orbital models, the properties of the pairing operators
can be studied under a mean-field approximation. We need to remember that the
five-orbital model is defined in terms of a pseudocrystal momentum ${\bf k}$ in
an extended Brillouin zone.\cite{graser} In terms of the real momentum, the 
unit cell of the Fe-As planes contains two Fe ions and, thus, the band 
structure 
is composed of 10 bands in the reduced, or folded, Brillouin zone (BZ). This
can be observed in Fig.~\ref{akwv0spm}(a) where the spectral function $A({\bf k},\omega)$
is shown for the non-interacting case along high symmetry directions in the 
folded BZ where the 10 bands are clearly seen. 
Thus, for each real momentum ${\bf q}$ there are 10 bands. Five of them are the
bands of the five-orbital model for the pseudocrystal momentum ${\bf k=q}$ and the
other five are obtained from the same model by setting the pseudocrystal momentum
to ${\bf k=q+Q}$, where ${\bf Q}=(\pi,\pi)$. Since in terms of the real 
momentum,
the basis of the five-orbital model is expanded by states with momentum ${\bf q}$
for orbitals $xz$ and $yz$ and momentum ${\bf q+Q}$ for the other three 
orbitals, this fact needs to be taken into account when pairing operators are
being constructed.
If only intraorbital pairing operators are considered
it is sufficient to build a $10 \times 10$ Bogoliuvov-deGennes (BdG) matrix, 
but for interorbital pairing between electrons in orbitals 1, 2 with electrons 
in orbitals 3, 4, or 5, it is necessary to construct a $20 \times 20$ BdG 
matrix since the $10 \times 10$ matrix only allows to consider intraorbital 
pairs with pseudocrystal momentum ${\bf Q}$ rather than $0$.\cite{foot1}


The $10 \times 10$ BdG Hamiltonian is given by:
\begin{equation}
H_{\rm BdG}=\sum_{{\bf k}}\Psi^{\dagger}_{\bf k}H^{\rm MF}_{\bf k}\Psi_{\bf k},
\label{13}
\end{equation}
\noindent with the definitions
\begin{eqnarray}
\Psi^{\dagger}_{\bf k}=(d^{\dagger}_{{\bf k},1,\uparrow},d^{\dagger}_{{\bf k},2,\uparrow},d^{\dagger}_{{\bf k},3,\uparrow},d^{\dagger}_{{\bf k},4,\uparrow},
d^{\dagger}_{{\bf k},5,\uparrow},\nonumber\\
d_{-{\bf k},1,\downarrow},d_{-{\bf k},2,\downarrow},d_{-{\bf k},3,\downarrow}d_{-{\bf k},4,\downarrow},d_{-{\bf k},5,\downarrow}),
\label{14}
\end{eqnarray}
\noindent and
\begin{equation}
H^{\rm MF}_{\bf k}=
 \left(\begin{array}{cc}
H_{\rm TB}({\bf k}) & P( {\bf k}) \\
P^{\dagger}({\bf k}) & -H_{\rm TB}({\bf k})
\end{array} \right),
\label{15}
\end{equation}
where each element represents a $5\times 5$ block with
$H_{\rm TB}({\bf k})$ given by Eq.~(\ref{1})
and
\begin{equation}
P({\bf k})_{\alpha,\beta}=Vf({\bf k})(\lambda_i)_{\alpha,\beta},
\label{16}
\end{equation}
\noindent where 
$V=V_0\Delta$ is determined by the product of an unknown pairing strength $V_0$
and a parameter $\Delta$ that arises from minimizing the mean-field 
equations, as already explained in previous literature.\cite{ib}

\subsection{Intraorbital Pairing Operators}

\subsubsection{Purely Intraband Pairing Operators}

The first issue we want to address is what is the form of the pairing 
operators resulting from purely intraband pairing interactions. The motivation 
is given by the fact that in standard BCS theory,  the
pairing occurs between particles with momentum 
equal in magnitude but opposite direction at a {\it common} Fermi surface. 
For this reason, even in multiorbital systems, it 
has been expected that pairing attraction should involve particles in the same 
band.\cite{suhl}
To determine whether a pairing operator consists of purely intraband 
matrix elements, we need to 
transform 
Eq.~\ref{15} from the orbital to the band representation in which $H_{\rm TB}$ is
diagonal. The transformation is given by 
$H_{\rm Band}({\bf k})=U^{\dagger}({\bf k})H_{\rm TB}({\bf k})U({\bf k})$,
where $U({\bf k})$ is the unitary change of basis matrix and 
$U^{\dagger}({\bf k})$ is the 
transpose conjugate of $U({\bf k})$. 
Since $U$ is unitary, for each value of ${\bf k}$,
$\sum_i(U_{i,j})^*U_{i,m}=\sum_i(U_{j,i})^*U_{m,i}=\delta_{j,m}$. 
Then, $H'_{\rm MF}=G^{\dagger}H_{\rm MF}G$ where 
$G$ is the $10\times 10$ unitary matrix composed of two $5\times 5$ blocks 
given by $U$. Then,
\begin{equation}
H'^{\rm MF}_{\bf k}=
 \left(\begin{array}{cc}
H_{\rm Band}({\bf k}) & P_{\rm B}( {\bf k}) \\
P_{\rm B}^{\dagger}({\bf k}) & -H_{\rm Band}({\bf k})
\end{array} \right),
\label{15a}
\end{equation}
with
\begin{equation}
P_{\rm B}({\bf k})=U^{-1}({\bf k})P({\bf k})U({\bf k}).
\label{15b}
\end{equation}
The most general form of a purely intraband pairing is given by
\begin{widetext}
\begin{equation}
H'^{\rm MF}_{\bf k}=\hspace{-0.5em}
 \left(\hspace{-0.5em}\begin{array}{cccccccccc}
\epsilon_1({\bf k}) & 0&0&0&0&\Delta_1( {\bf k})&0&0&0&0 \\
0&\epsilon_2({\bf k}) &0&0&0&0&\Delta_2( {\bf k})&0&0&0 \\
0 & 0&\epsilon_3({\bf k})&0&0&0&0&\Delta_3( {\bf k})&0&0 \\
0&0&0&\epsilon_4({\bf k}) &0&0&0&0&\Delta_4( {\bf k})&0 \\
0 & 0&0&0&\epsilon_5({\bf k})&0&0&0&0&\Delta_5( {\bf k}) \\
\Delta^*_1( {\bf k})& 0&0&0&0&-\epsilon_1({\bf k}) &0&0&0&0\\
0&\Delta^*_2( {\bf k}) &0&0&0&0&-\epsilon_2({\bf k}) &0&0&0 \\
0& 0&\Delta^*_3( {\bf k}) &0&0&0&0&-\epsilon_3({\bf k})&0&0\\ 
0&0&0&\Delta^*_4( {\bf k}) &0&0&0&0&-\epsilon_4({\bf k}) &0 \\
0& 0&0&0&\Delta^*_5( {\bf k}) &0&0&0&0&-\epsilon_5({\bf k})
\end{array} \hspace{-0.5em}\right),
\label{17}
\end{equation}
\end{widetext}
where $\epsilon_j({\bf k})$ are the eigenvalues of $H_{\rm TB}({\bf k})$, 
while $\Delta_j({\bf k})$ denotes the band and momentum dependent pairing
interactions. Notice that if $\lambda_i=\lambda_0$ in Eq.(\ref{16}), 
$P({\bf k})$ in Eq.(\ref{15b}) is proportional to the identity matrix and,
thus, $P_{\rm B}({\bf k})$ is diagonal with $\Delta_j( {\bf k})=\Delta( {\bf k})=
Vf( {\bf k})$ for all $j$. This indicates that the {\it intraorbital} 
operator that
pairs electrons in each orbital with {\it equal strength}, gives rise to 
a purely
intraband pairing operator where electrons in each band are subject to
an identical pairing attraction. 
In this case, diagonalizing Eq.(\ref{17}) we obtain
$E_j({\bf k})=\sqrt{\epsilon_j({\bf k})^2+|\Delta({\bf k})|^2}$. 
Then, at ${\bf k}_F^j$
where $\epsilon_j({\bf k}_F^j)=0$ we obtain 
$E_j({\bf k}_F^j)=|\Delta({\bf k}_F^j)|$.
This means that the superconducting gaps at the
FS determined by different bands must satisfy that 
$\Delta'({\bf k}'_F)/\Delta({\bf k}_F)=
\frac{f( {\bf k}'_F)}{{f( {\bf k}_F)}}$ where ${{\bf k}'_F}$ and ${{\bf k}_F}$ represent
the Fermi momentum of two FS defined by two different bands. In the case of 
the standard low-temperature 
BCS pairing $f( {\bf k})=1$, implying that momentum independent gaps of 
equal
magnitude should open in all the FSs. However, for the pnictides it is 
believed that a non-BCS interaction provides the source of 
pairing.\cite{phonon0} Then, we must consider the case in which 
$f({\bf k})\ne 1$. Notice that since $\lambda_0$ transforms according 
to $A_{\rm 1g}$, the symmetry of this purely intraband pairing operator is 
then totally determined by the symmetry of $f({\bf k})$.

\subsubsection{Single Gap: The $S \pm$ pairing operator}

In the particular case in which $f({\bf k})=\cos k_x \cos k_y$,
$P({\bf k})$ transforms according to $A_{\rm 1g}$ and it represents the simplest 
form of the well 
known $S \pm$ pairing state,\cite{kuroki,Mazin:2008p1695,korshunov,parker,seo} 
which
will be characterized by gaps of approximately the same magnitude in the 
electron and hole pockets if there is good $(\pi,0)$ and $(0,\pi)$ 
nesting. Note that the gap on 
the two hole FSs would differ only if they have considerably different Fermi 
momentum which is not the case in the pnictides since both hole pockets are
very close to each other. In the five-orbital model we found that 
$r=\frac{f( {\bf k'})}{{f( {\bf k})}}\sim 1$ for the external hole and 
electron pockets, while $r\sim 0.9$ for the external and internal hole pockets.
This means that the gaps in the two hole pockets would have a difference of 
only about 
10\%. The two Fermi surfaces would have to be much more separated from 
each other in order to develop the experimentally observed 
difference of 50\% reported via ARPES experiments.\cite{arpes2}

\begin{figure}[thbp]
\begin{center}
\includegraphics[width=7cm,clip,angle=0]{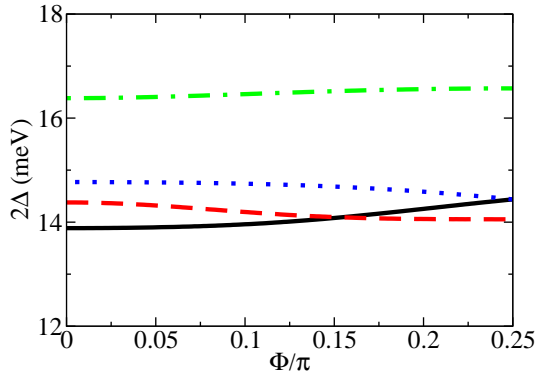}
\vskip 0.3cm
\caption{(Color online) The gap at the internal-hole (dashed-dotted line), 
external-hole (dashed line), external-electron (continuous line) and
internal-electron (dotted line) Fermi surfaces 
for the $S \pm$ pairing operator with $V=0.02$. 
Results are
shown as a function of the angle $\Phi$ between 0 and $\pi/4$ 
measured with respect to the $k_x$ axis in  counterclockwise [clockwise] 
direction for the hole [electron (at X)]-pockets. The results are in the 
folded Brillouin zone.\label{spmgap}}
\end{center}
\end{figure}

In Fig.~\ref{spmgap} we show the gap as a function of the angular
position (from the $x$ to the $y$ axis) along the four Fermi surfaces (two 
electron and two hole pockets) for 
the five-orbital model in the folded BZ. Note that the values of the gap beyond
$\Phi=\pi/4$ can be obtained by symmetry from the values shown in Fig.~\ref{spmgap}.
The interaction $V=0.02$ was chosen in order to obtain a
quantitative match with experimentally reported values of the gaps that 
range between 1 to 20 meV.\cite{rossle}
As expected, it can be seen that the gaps at the external hole and electron 
pockets (dashed, continuous, and dotted lines) almost overlap with each other 
and have a magnitude of about 7 meV while at 
the internal hole pocket (dotted-dashed line) the gap is around 8 meV, i.e.
only about 10\% larger. Note that the momentum dependence observed here likely will
be negligible when
considering experimental uncertainties. Thus, these results would be in good agreement
with the ARPES measurements reported in Ref.~\onlinecite{arpes2} if 
we assume that
the 50\% smaller gap arises in a third hole pocket, rich in $x^2-y^2$ orbital 
content, that is not present in this five-orbital model.

The mean-field calculated spectral functions are shown in 
Fig.~\ref{akwv0spm}(b) along high symmetry directions in the reduced BZ. In the 
figure, $V$=$0.2$ is used \cite{com} and it can be seen 
that the gaps, that are very similar in 
magnitude, have 
opened on all the original FSs. As expected, a numerical inspection of the 
whole BZ shows that there are no nodes. It can be observed that band 
distortions and ``shadow'' (or Bogoliubov) band spectral weight \cite{shadow} 
develop 
only in a small region 
around the chemical potential.

If a different form factor such as $f({\bf k})=\cos k_x -\cos k_y$ is 
considered, the pairing operator would still be purely intraband but with 
symmetry $B_{\rm 1g}$. While now nodes would appear on the FS because the pairing
interaction vanishes for $k_x=\pm k_y$, the gaps on the different 
FSs still only will differ by the ratio of the form factors at the 
respective Fermi momenta. 


\subsubsection{Multiple Gaps}

Since several experimental\cite{szabo,arpes2,arpes22,matano,deve} and 
theoretical\cite{tesa,boyd} efforts have reported the existence
of at least two independent gaps at the Fermi surfaces of some pnictides,
the next issue to consider is whether there is any other pairing state allowed 
by symmetry that is purely intraband and able to generate independent gaps
in at least one of the FSs. It can be shown that, with the 
exception of $\lambda_0$, all the other 24 $\lambda_i$ matrices become 
non-diagonal in the band representation, i.e., $\lambda_i^B=U^{-1}({\bf k})
\lambda_i U({\bf k})$ is non-diagonal at least for some values of ${\bf k}$ 
in the BZ. This is true even for $\lambda_i$ matrices that
are diagonal in the orbital representation such as $\lambda_3$, $\lambda_8$,
$\lambda_{23}$, and $\lambda_{24}$. Thus, {\it all} these purely intraorbital 
pairing operators lead to {\it interband} matrix elements in their band 
representation. Note that to generate an 
intraorbital  pairing operator that couples electrons in different orbitals 
with arbitrary strengths we need to consider linear combinations of these 
diagonal pairing matrices using form factors that provide a well defined
symmetry, since we assumed, guided by numerical simulations in models with two 
and three orbitals, that the pairing operator connects non-degenerate ground 
states that transform according to one of the one-dimensional irreducible 
representations of $D_{\rm 4h}$.

We will consider a linear combination of
the four intraorbital pairing operators allowed by symmetry that do not 
require the pairing to be the same in all orbitals. $\lambda_0$ is excluded 
because we already know that it does not produce interband pairing and, thus, 
it will not contribute to off-diagonal elements in the band representation of 
the pairing operator.
The linear combination to be considered is given by:
\begin{equation}
\Lambda=\alpha_1\lambda_8 f_1({\bf k})+\alpha_2\lambda_{23} f_2({\bf k})
+\alpha_3\lambda_{24} f_3({\bf k})+\alpha_4\lambda_3 f_4({\bf k}),
\label{20}
\end{equation}
\noindent where $f_i({\bf k})$ are form factors chosen such that all the terms
in $\Lambda$ transform according to the same irreducible representation. 
Notice that $\lambda_8$, $\lambda_{23}$, and $\lambda_{24}$ transform 
according to
$A_{\rm 1g}$, but $\lambda_3$ transforms according to $B_{\rm 1g}$. 
Thus, for example, a nearest-neighbor pairing operator that 
transforms according to $A_{\rm 1g}$ will be 
obtained by setting $f_1({\bf k})=f_2({\bf k})=f_3({\bf k})=\cos k_x+\cos k_y$
and $f_4({\bf k})=\cos k_x-\cos k_y$. In the band representation $\Lambda$
becomes
\begin{equation}
\Lambda_b=U^{-1}({\bf k})\Lambda U({\bf k}).
\label{21}
\end{equation}
If the pairing is purely intraband then $\Lambda_b$ has to be diagonal. But 
from the orthogonality properties of $U$ we see that a non-diagonal
element $(\Lambda_b)_{i,j}$ can only be zero if $\Lambda_{i,i}$ is the same
for all $i$. However, this is only true for the pairing operator proportional 
to $\lambda_0$ and we have proven numerically that only for $\alpha_j=0$ in
Eq.~\ref{20} is $\Lambda_b(m,n)=0$ for $m\ne n$. Thus, intraorbital pairing 
with orbital dependent 
strength leads to interband components of the pairing interaction.\cite{foot} 

Then, we conclude that if the pairing interaction is purely intraband the 
pairing mechanism should be associated to a degree of freedom that couples to 
the five different orbitals with the \emph{same} strength. 
This means that the gaps in 
all different FSs must be related, i.e., the symmetry does not allow 
unrelated gaps in this case. Conversely, if independent gaps are observed 
on different FSs the symmetry of the highly hybridized orbitals indicates that
interband interactions would be present at least in some regions of momentum 
space, as it will be discussed below.\cite{foot3}

\begin{figure}[thbp]
\begin{center}
\includegraphics[width=8cm,clip,angle=0]{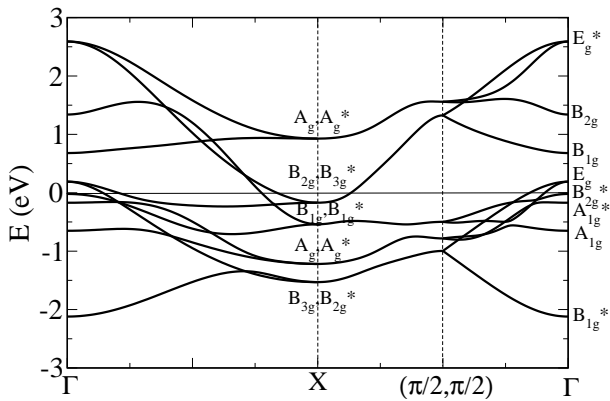}
\vskip 0.3cm
\caption{Band structure of the five-orbital model along high 
symmetry directions in the reduced Brillouin zone. The irreducible 
representations of $D_{\rm 4h}(\Gamma)$ and $D_{\rm 2h}(X)$, characterizing 
the bands 
at high symmetry points in the reduced BZ, are indicated. For more details
see also the discussion in Sec.~\ref{disc}. \label{bandas}}
\end{center}
\end{figure}
Some ARPES data \cite{arpes2} indicate a 
superconducting state with momentum independent gaps with value $\Delta$ at 
the hole and electron
pockets and $\Delta/2$ at a third hole pocket not present in the 
five-orbital model, at least  
with the parameters used here and in Ref.~\onlinecite{graser}. 
However, a slight modification of the parameters, without affecting the 
symmetry of the Hamiltonian,\cite{spanish} would create an additional hole 
pocket around 
$\Gamma$ of $x^2-y^2$ character. This is the state with symmetry
$B_{\rm 2g}^*$ at the $\Gamma$ point shown in 
Fig.~\ref{bandas}.  According to this figure,
close to the hole FSs around $\Gamma$ the pairing
matrix should be diagonal with elements $\Delta$ ($\Delta/2$) for the bands 
labeled $E_{\rm g}$ ($B_{\rm 2g}^*$) at $\Gamma$, 
which in the unfolded BZ corresponds to {\it (i)}
a diagonal pairing matrix at ${\bf k}\approx{\bf k}_F^h$ 
($h$ denoting the hole Fermi surface) with at least two diagonal elements 
equal to $\Delta$, and {\it (ii)} a diagonal pairing matrix at 
${\bf k}\approx{\bf k}_F^h+{\bf Q}$ with at least
one diagonal element equal to $\Delta/2$. As just discussed, these 
two different values of a momentum-independent gap cannot arise from
purely intraband pairing interactions in a highly hybridized system.

\subsubsection{Nodeless Gaps with Interband Pairing Matrix Elements}

Since several experimental ARPES studies of the pnictides 
appear to indicate that 
nodeless gaps open at all the FSs in the superconducting 
state,\cite{arpes,arpes2,arpes3,arpes22,hsieh} we will now identify 
the pairing operators that produce nodeless gaps that are 
allowed by the symmetries of the five-orbital model. 

In a previous study of a  three-orbital model we found that, in addition to the
$S \pm$ pairing operator, there was another nodeless pairing operator with 
both intra and 
interband matrix elements that was called $S_{\rm IB}$.\cite{our3} 
It leads to intraorbital pairing with different strength for the orbitals 
$xz$/$yz$ and $xy$, and it transforms according
to $A_{\rm 1g}$. In the five-orbital model a similar result has been obtained. 
In fact, we have found several linear combinations 
with $A_{\rm 1g}$ symmetry of the intraorbital pairing matrices $\lambda_i$ 
with $i=0$, 8, 23, and 24 that provide
nodeless gaps. In these pairing states, $f({\bf k})=1$ or $\cos k_x\cos k_y$
and the pairing interaction is \emph{not} the 
same in  the different orbitals. The pairing matrix has the form
\begin{equation}
P({\bf k})=f({\bf k})
 \left(\begin{array}{ccccc}
V_1 & 0 & 0 & 0 & 0  \\
0     & V_1 & 0 & 0 & 0  \\
0 & 0 & V_2 &  0 & 0  \\
0 & 0 & 0 & V_3 &  0  \\
0 & 0 & 0 & 0 & V_4 
\end{array} \right),
\label{pairing1}
\end{equation}
\noindent where $V_i$ denote the different pairing strengths. Examples 
of parameter sets for 
which nodeless gaps are found are
$(V_1,V_1,V_2,V_3,V_4)=V(1,1,0,2,0)$, 
$V(1,1,0,1,0)$, and $V(1,1,0,0,0)$. These operators pair electrons
in the orbitals that contribute the most to the FS, but it is important to 
notice that they are {\it not diagonal} in the band representation. 

\begin{figure}
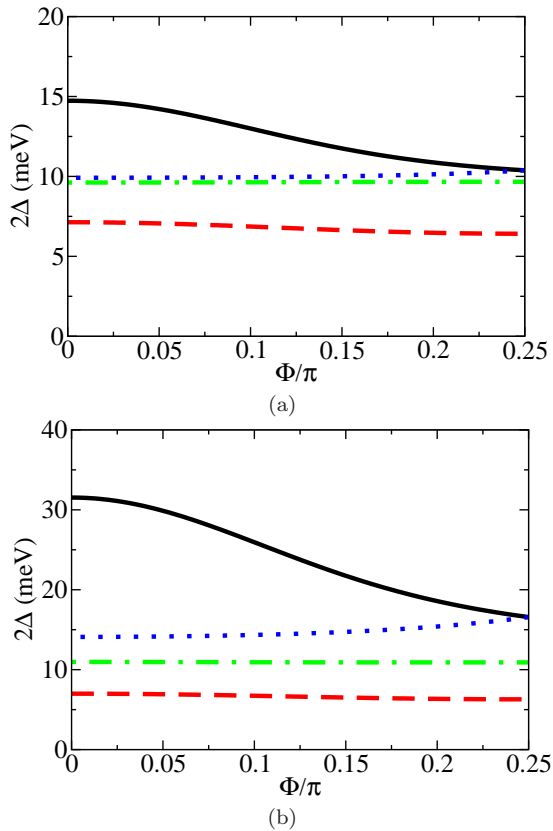

\subfigure[]{\includegraphics[width=0.4\textwidth,clip] {gapSib3_4a}
\label{gv01}}\\[-.5em]
\subfigure[]{\includegraphics[width=0.4\textwidth,clip]
{gapexp_4b}\label{gexp}}
\caption{(Color online) The gap at the internal-hole (dashed-dotted line), 
external-hole (dashed line), external-electron (continuous line), 
and internal-electron (dotted line) Fermi surfaces 
for the nodeless pairing operator $S_{\rm IB}$ with
(a) $(V_1,V_1,V_2,V_3,V_4)=(0.01,0.01,0.0,0.02,0.0)/\sqrt{6}$ and (b) 
$(V_1,V_1,V_2,V_3,V_4)=(0.005,0.005,0.0,0.02,0.0)$. 
Results are
shown as a function of the angle $\Phi$ between 0 and $\pi/4$ 
measured with respect to the $k_x$ axis in  counterclockwise (clockwise) 
direction for the hole [electron (at X)]-pockets.
The notation is as in
Fig.~\ref{spmgap}.\label{gap}}
\end{figure}

The momentum dependence of the gaps at the FS is shown in Fig.~\ref{gv01}
for the $S_{\rm IB}$ pairing operator with $f({\bf k})=\cos k_x\cos k_y$ and
$(V_1,V_1,V_2,V_3,V_4)=(0.01,0.01,0.0,0.02,0.0)/\sqrt{6}$, as special case. 
To reproduce experimental values for the 
gap\cite{rossle} $V=0.01/\sqrt{6}$ has been 
chosen. It can be observed that in this case the gaps for 
the internal hole and electron pockets in panel (a)
(dashed-dotted and dotted lines) have a negligible momentum dependence 
and an average value of about 5 meV. The external hole pocket has a smaller
gap, about 3.5 meV, with a very weak momentum dependence (dashed line). The 
strongest dependence with momentum occurs in the external electron pocket 
(continuous line) with the gap
reaching a value of about 7.5 meV at the point along the $x$-axis where the 
electron pocket is the closest to the hole pockets and has pure $xy$ orbital 
character. As the point where the two electron-like FSs intersect each other 
($\Phi=\pi/4$) is approached, the external electron-pocket acquires $xz/yz$ 
character and the magnitude of the gap converges to the almost momentum 
independent value of 5 meV that characterizes the internal electron-pocket. 
Thus, this is a 
case in which a direct measurement of the gaps would indicate the presence of 
three ``independent'' gaps. The gaps in the internal electron and hole pockets 
behave as expected in the $S \pm$ pairing 
scenario,\cite{kuroki,Mazin:2008p1695,korshunov,parker,seo} while the external
hole and electron pockets show a different gap value. 
While this would agree with 
experimental results that favor multiple 
gaps\cite{szabo,arpes2,arpes22,matano,deve,tesa,boyd} it is important to
realize that this pairing operator contains interband matrix elements that
must be incorporated in the description of a possible pairing 
mechanism.

Modifying $V_i$ we can tune the relative gap values, obtaining
three almost momentum independent gaps with values 3.5 meV, 5.5 meV, and
7.5 meV (in the external-hole, internal-hole, 
and internal-electron pockets, respectively)
as shown in Fig.~\ref{gexp}. In addition,
a gap with momentum dependence ranging between 15 meV and 7.5 meV 
appears at the external electron pocket which has the largest $xy$ 
contribution. Thus, this interaction would lead to four apparently independent 
nodeless gaps.

\subsection{Interorbital Pairing Operators}

As discussed above,  due to the strong hybridization of 
all five $d$ orbitals we have verified that all the intraorbital pairing
operators allowed by the lattice and orbital symmetries of the pnictides lead
to interband pairing interactions if $\lambda_i\ne\lambda_0$. 
Interband pairing has always been considered unlikely in BCS theory
\cite{suhl} because the pairing attraction acts in a very narrow energy range 
around the FS. However, 
in a system in which two FSs formed by different bands are 
very close to each other (e.g. 
the two hole pockets in the pnictides that cannot 
be distinguished in ARPES experiments,\cite{arpes2,arpes22,hsieh} or the two 
electron pockets that intersect at two points) then the formation of interband 
pairs, as it was described in Ref.~\onlinecite{ib}, could occur. In addition, 
numerical Lanczos
studies of a two-orbital model for the pnictides suggest that an interorbital
pairing state with $B_{\rm 2g}$ symmetry, with interband components, could 
be the favored pairing state in the intermediate Hubbard $U$ 
regime.\cite{Daghofer:2009p1970,moreo,rong}
For this reason, here results will be presented for some interorbital
pairing operators 
that have interband attraction, at least in some regions of the BZ. The 
effects of this interaction will be addressed later in Sec.~\ref{disc}.
\begin{figure}[thbp]
\begin{center}
\includegraphics[width=7cm,clip,angle=0]{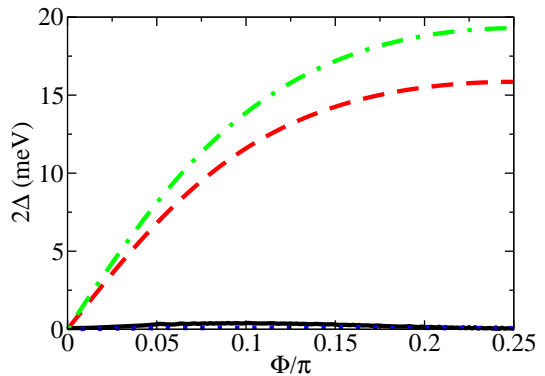}
\vskip 0.3cm
\caption{(Color online) The gap at the internal-hole (dashed-dotted line),
external-hole (dashed line), external-electron (continuous line), and
internal-electron (dotted line) Fermi surfaces 
for the interorbital pairing operator 
$B_{\rm 2g}$ with $V=0.012/\sqrt{5}$. 
Results are
shown as a function of the angle $\Phi$ between 0 and $\pi/4$ 
measured with respect to the $k_x$ axis in  counterclockwise [clockwise] 
direction for the hole [electron (at X)]-pockets.
The notation is as in Fig.~\ref{spmgap}\label{gapB2g}}.
\end{center}
\end{figure}

First let us consider the $B_{\rm 2g}$ pairing 
operator mentioned in the previous 
paragraph. It is given by
\begin{equation}
\Delta^{\dagger}_{B_{\rm 2g}}({\bf k})=V(\cos k_x+\cos k_y)
\sum_{\alpha\ne\beta=1}^2 d^{\dagger}_{{\bf k},\alpha,\uparrow}
d^{\dagger}_{-{\bf k},\beta,\downarrow}.
\label{b2g}
\end{equation}

\begin{figure}[thbp]
\begin{center}
\includegraphics[width=8cm,trim = 30 30 25 35,clip,angle=0]{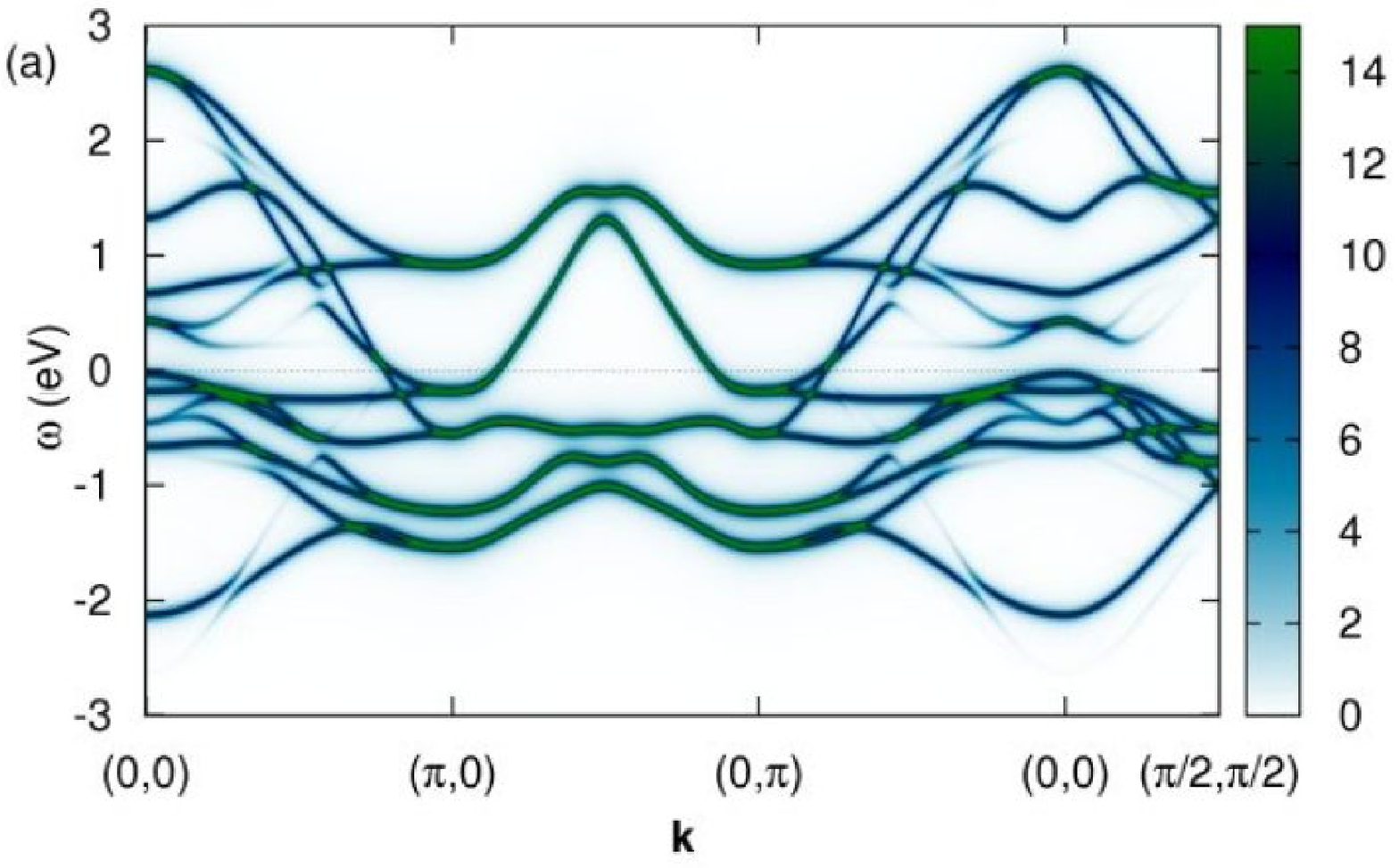}
\includegraphics[width=8cm,trim = 30 30 25 35,clip,angle=0]{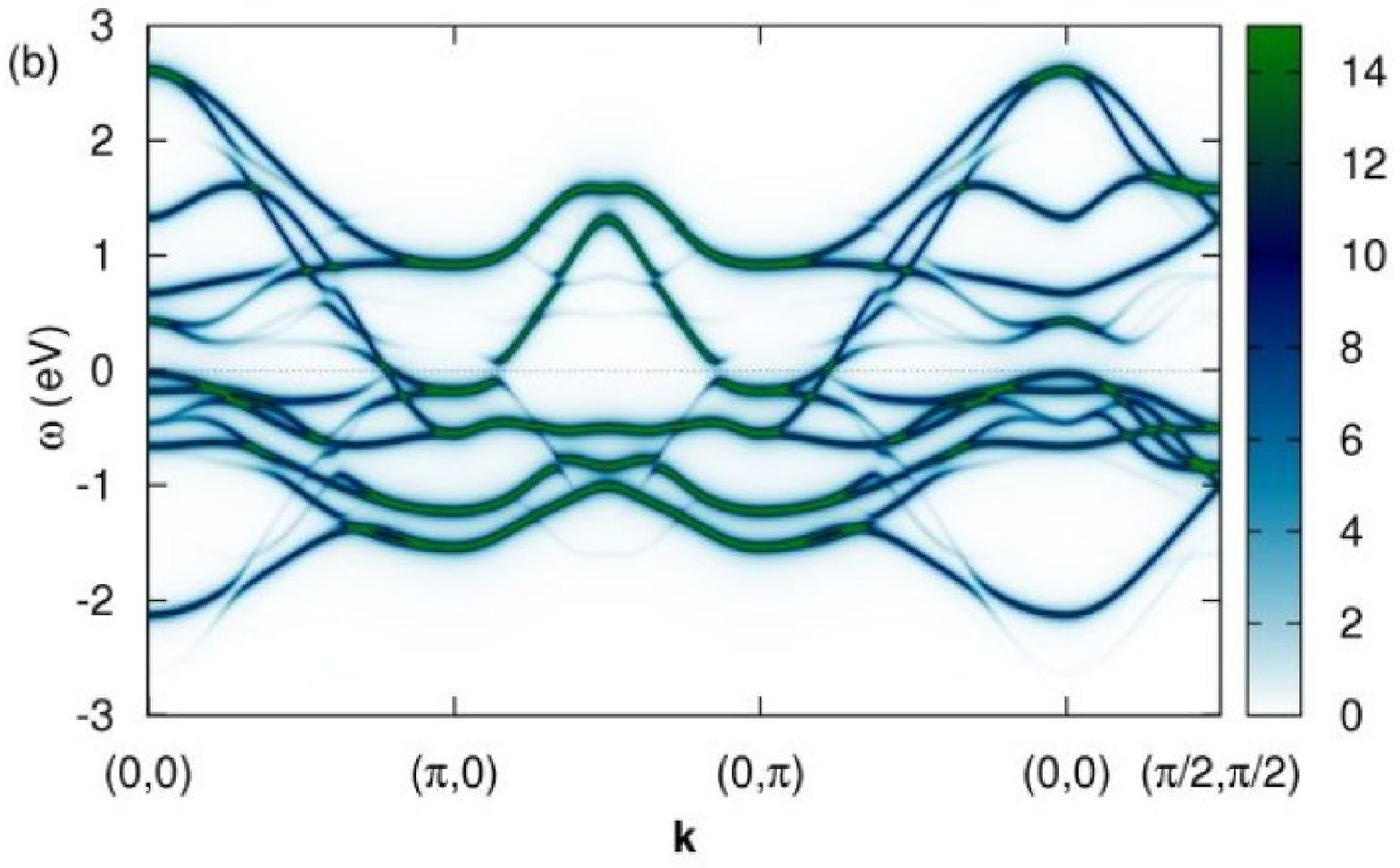}
\vskip 0.3cm
\caption{(Color online) The intensity of the points represent the values of the
spectral function $A({\bf{k},\omega})$ for the five-orbital model with pairing
interaction  $V=0.2$, for the pairing operator (a) $B_{\rm 2g}$
and (b) extended $B_{\rm 2g}$
discussed in the text. The results are shown in the reduced BZ.}
\label{IVVg}
\end{center}
\end{figure}

For this operator, the structure of the gap strongly depends on the value of the
pairing attraction $V$. For values of $V$ compatible with the order of 
magnitude of the gaps reported from experiments in the 
pnictides, which are
of the order 
of meV, the behavior of the gap at the different FSs is shown in 
Fig.~\ref{gapB2g}. The gap on the two hole pockets presents nodes along the $x$ 
($\Phi=0$) and $y$ ($\Phi=\pi/2$) axes (the latter not shown explicitly 
since the results are 
symmetric and thus $\Delta(\Phi)=\Delta(\pi/2-\Phi)$). Both gaps are maximized 
along the diagonal direction ($\Phi=\pi/4$). Thus, it would resemble 
``$d_{xy}$-wave'' behavior in experiments where only the node location but not 
the phase of the gap can be measured. The reason for the existence of the 
nodes is that along the $x$ and $y$ axes the pairing interaction is purely 
interorbital, but $V$ is not strong enough to open a gap (a finite value of 
$V$ is needed to open a gap with purely interorbital pairing, as
discussed in previous literature~\cite{ib}) and 
that effect creates the node. By contrast, in Fig.~\ref{IVVg}(a) we show the mean-field 
calculated spectral 
functions along some high symmetry directions of the reduced BZ for the same 
pairing operator but with a larger $V=0.2$. 
In this case, the gaps have opened on the hole-pocket FSs around 
the $\Gamma$ point. But for this value of $V$ the gap is of the order of 
100 meV, i.e. too large compared with the experimental results for the pnictides.

Another characteristic of the $B_{\rm 2g}$ pairing operator is that 
the gap in the electron pockets presents nodes along the $x$ and $y$ axes (for the 
same reason than the hole pockets 
at weak $V$) but also at the points where the two electron pockets cross, 
$\Phi=\pi/4$, because $f({\bf k})=0$ there. For $V<0.01$ the gap is much smaller than
the one on the hole-pockets, as shown in Fig.~\ref{gapB2g}. 
As $V$ increases, a gap opens along the $x$ and $y$ axes 
for the internal electron pocket, as it can be observed in Fig.~\ref{IVVg}(a)
for $V=0.2$. However, in this case nodes remain along these axes for the 
external electron pocket. The reason is that at
these points the FS arises from bands that have mostly character
$xz/xy$ or $yz/xy$\cite{graser} and, thus, the operator 
$\Delta^{\dagger}_{B_{2g}}$, which couples electrons in the $xz/yz$ orbitals, 
is not effective at opening a gap along the momentum axis where one of 
the FSs has purely $xy$ character. 
While this gap structure for the $B_{\rm 2g}$ pairing state disagrees with ARPES 
measurements it is important to keep in mind that the existence of 
nodes in some pnictides has 
been reported by several groups using other experimental techniques. Thus, it is still possible
that the surface that is actually tested by ARPES does not present the same behavior 
as the bulk in the pnictides.

A natural generalization of the operator $\Delta^{\dagger}_{B_{\rm 2g}}$ 
to include 
the $xy$ orbital was 
provided in Ref.~\onlinecite{our3}. The extended operator
is a linear combination of $\Delta^{\dagger}_{B_{\rm 2g}}$ (Eq.~\ref{b2g}) and
\begin{eqnarray}
\Delta^{\dagger}_{V_g}({\bf k})=V'[\sin k_x
\sum_{\alpha\ne\beta=2,4} d^{\dagger}_{{\bf k},\alpha,\uparrow}
d^{\dagger}_{-{\bf k},\beta,\downarrow}-\notag\\
\sin k_y
\sum_{\alpha\ne\beta=1,4} d^{\dagger}_{{\bf k},\alpha,\uparrow}
d^{\dagger}_{-{\bf k},\beta,\downarrow}].
\label{Vg}
\end{eqnarray}
The total pairing operator has symmetry $B_{\rm 2g}$. Note that now
the $10 \times 10$ matrix given in Eq.~\ref{13} will provide pairs with
pseudocrystal momentum 0 for the particles in orbitals 1 and 2, but 
${\bf Q}$ for one of the particles in orbital 4 and the other in 
1 or 2. Thus, we must
consider the $20 \times 20$ BdG matrix mentioned earlier and we have studied
the extended $B_{\rm 2g}$ pairing with the two possible pseudocrystal momenta
0 and ${\bf Q}$. This generalized form removes the nodes at the crossing point
of the two electron pockets, since $\Delta^{\dagger}_{V_g}$ is finite at those
points. 

In Fig.~\ref{IVVg}(b) we show the spectral
function for $V=V'=0.2$ for pairs with 0 pseudocrystal momentum. 
It can be seen that isolated nodes at the electron pockets remain only
at the points along the momentum axis where they have purely $xy$ character. 
However, for smaller values of $V=V'$ that produce gaps in the 
meV range, we have found that nodes along the axis remain and the
gaps resemble very much those obtained with the original $B_{\rm 2g}$ pairing 
operator shown in Fig.~\ref{gapB2g}.

\section{Discussion}\label{disc}

In theoretical studies of multiorbital systems it has been 
``traditional'' to consider  
non-hybridized orbitals/bands.\cite{suhl} While this approach works for simple 
cases, in the pnictides the hybridization of the orbitals is 
strong.\cite{phonon0,first} As a result, the energies in the band
representation $\epsilon_j({\bf k})$ have several accidental degeneracies within 
the Brillouin Zone which means that the bands cross at several 
points and are, thus, very 
entangled.\cite{foot2}
This band entanglement is apparent in 
Fig.~\ref{bandas}, where the eigenvalues of $H_{\rm TB}$ are shown along 
high symmetry directions in the folded BZ.
The irreducible representations characterizing the bands at some high symmetry 
points (such as $\Gamma$ with symmetry $D_{\rm 4h}$ 
and $X$ with symmetry $D_{\rm 2h}$) 
are indicated, and the star labels bands determined by states with 
pseudocrystal momentum ${\bf k}+{\bf Q}$. For example, the
band labeled $B_{\rm 1g}$ at  $\Gamma$, crosses two other bands 
along the $\Gamma-X$ direction. Thus, the 
approach used in early studies
by Suhl {\it et al.}, where a gap $\Delta_j$ is associated to a
band with energy $\epsilon_j({\bf k})$ in all the BZ, becomes ambiguous in this 
case. In a system with strongly hybridized orbitals, it is more 
reasonable to define pairing operators in the orbital representation because the
orbital basis is globally well-defined at all points in the BZ, while the band 
assignation is local. In addition,  as it was shown in the previous Section, 
the orbital representation provides
the natural framework to construct the pairing operators that are 
allowed by symmetry.
All the active degrees of freedom need to 
be incorporated 
in the pairing operators since otherwise the results can be misleading, as in 
the early studies of magnetization when the contribution of the spin to the 
spatial wave function was disregarded. Then, when
intraband pairing operators are constructed their orbital content should be provided.

Another important point is to understand the consequences of interband matrix
elements 
arising from the symmetry of the pairing operators. As shown in 
Ref.~\onlinecite{ib}, at points in the BZ where there are intra and interband
elements in $P_{\rm B}({\bf k})$ the pairs will be formed by electrons in the same
band if the pairing interaction is weak or intermediate, which seems to be the case in the 
pnictides. It is only at the very few points on 
the FS where $P_{\rm B}({\bf k})$ has finite (zero) non-diagonal (diagonal) 
elements that the 
pairing attraction will be purely interband. At these points, if the pairing 
attraction is weak we would expect to observe nodes.\cite{ib} 
It is only when the two FS
are very close (as for the two almost degenerate hole pockets), and when the pairing
interaction is strong enough, that interband pairs would be possible.\cite{ib}


As pointed out in the Introduction, the properties of the pairing operator in
the pnictides have not been established yet and data obtained with different 
experimental techniques are in disagreement. ARPES results are interpreted as
indicative of nodeless gaps with at most a very weak momentum dependence. Some
groups have identified gaps with the same magnitude in the electron and one of 
the hole pockets,\cite{arpes,arpes2,arpes3,arpes22,hsieh} while others have 
found them to be different.\cite{rossle} In addition, the gap in a third
hole pocket, not of $xz$/$yz$ character, 
is found to be different from the gap in 
the electron pockets and, thus, the existence of two and, even three, gaps in 
these materials has been proposed. The symmetry arguments presented here 
indicate that truly momentum independent gaps would have to be equal 
(including the sign of the order parameter) in all
Fermi surfaces. If a momentum dependence is allowed, then a pairing 
state such as
the $S \pm$ could arise. In this case the gap in the hole and electron pockets
related by nesting should be very similar (although with opposite signs in the order
parameters) and 
the gap in the additional hole pocket would differ only by the ratio between
the form factors $f({\bf k})$ at the locations of the two different 
hole-pockets. Thus, there should be only one gap in the sense
that the coupling between the electrons and the interaction causing the pairing
will not be orbital dependent and the band dependence is only due to the 
different Fermi momenta.

We have also shown that a pairing interaction that couples with 
an orbital-dependent strength to the electrons leads to a 
pairing matrix that is not diagonal in the band representation. Some of these operators open weakly 
momentum dependent nodeless gaps in the different FSs. While this kind of 
pairing operator could agree with some experimental results, it would be 
necessary to take into account the interband pairing interactions when 
developing the associated microscopic pairing mechanism. 

Note that
the existence of nodes in the pnictides is 
in fact supported by transport experiments,\cite{nodal1,nodal2,nodal3,Ahilan,nakai,Grafe,Y.Wang,matano,mukuda,millo,wang-nodes} 
in clear contradiction with ARPES.
Thus, more experimental work is
needed in order to clarify this issue. We have found a large variety of
nodal pairing operators that respect the symmetry of the pnictides: this includes
those proportional to the identity in the orbital sector, with their nodes 
arising from the zeroes in $f({\bf k})$, and also other pairing operators
in which electrons in 
different orbitals are subject to different pairing strengths and, 
thus, give rise to interband terms in the pairing matrices.  

The $S \pm$ pairing operator with 
symmetry $A_{\rm 1g}$ appears to be
the favored one in the 
literature,\cite{kuroki,Mazin:2008p1695,korshunov,parker,seo} but numerical
calculations in a two-orbital model\cite{Daghofer:2009p1970,moreo} while 
indicating
that this pairing prevails in strong coupling, lead
in the intermediate regime to
a pairing state made of electrons in different orbitals and with symmetry 
$B_{\rm 2g}$. 
Since the orbitals $xz$ and $yz$ are strongly hybridized 
forming the bands that produce the hole pockets, this kind of pairing could 
be possible. If realized, it would induce nodes on the hole and electron 
pockets and the pair formation would be much stronger in the hole 
than in the electron pockets. Interestingly, the gaps in the two hole pockets
would be different (see Fig.~\ref{gapB2g}) which may be in agreement with some
of the experimental results that indicate two nodal gaps.\cite{heat}


\section{Conclusions}\label{concl}

Summarizing, we have shown that in a model that retains the symmetry of the 
FeAs planes for the pnictides and considers the five $d$ orbitals of the Fe 
ions, a purely intraband pairing operator can only result from an intraorbital
pairing interaction that affects electrons in the different orbitals with 
{\it identical strength}. In this case, the symmetry of the pairing operator is 
entirely determined by the form factor $f({\bf k})$ which depends only 
on the
spatial location of the particles that form the pairs. As a result, gaps in 
different portions of the FS can differ only by the ratios of the form 
factors, i.e., two or more unrelated gaps {\it cannot} occur. 
Conversely, multiple
gaps as observed experimentally, 
or orbital dependent pairing attractions, would indicate interband pairing
interactions at least in some regions of the BZ. Then, this feature should be 
incorporated in theoretical proposals for the pairing mechanism.  If there 
are special points in the BZ where the attraction is purely interband, then 
nodes or interband pairing will occur depending on whether the interaction is 
weak or strong. Experimental measurements of the gap magnitude indicate that
the pairing attraction is weak, thus nodes rather than interband pairing would
be expected.

The present analysis suggests that if the pairing mechanism is purely 
intraband, as assumed by many, then
no unrelated gaps should occur in the different portions of the FS. 
Reciprocally, if it is 
experimentally confirmed the existence of two or more unrelated gaps 
this would point to the need
to consider interband, in addition to intraband, pairing interactions in any 
realistic microscopic description of the pairing mechanism.

\section{Acknowledgments} This work was supported by the National Science
Foundation grant DMR-0706020 
and by the
Division of Materials Sciences and Engineering, Office of Basic Energy Sciences,
U.S. Department of Energy, M.~D. acknowledges partial suppost from the DFG under the Emmy-Noether program.

\appendix
\section{Expressions for the Tight Binding Hamiltonian}
\label{app:1_eff}

\begin{eqnarray}
\xi_{11}&=&2t_{x}^{11}\cos k_x +2t_{y}^{11}\cos k_y + 4t_{xy}^{11}
\cos k_x \cos k_y +\notag \\
&& 2t_{xx}^{11}(\cos 2k_x - \cos 2k_y) +
4t_{xxy}^{11} \cos 2k_x \cos k_y \notag \\
&& + 4t_{xyy}^{11}\cos 2k_y \cos k_x+4t_{xxyy}^{11}\cos 2k_x \cos 2k_y\\
\xi_{22}&=&2t_{y}^{11}\cos k_x +2t_{x}^{11}\cos k_y  + 4t_{xy}^{11}
\cos k_x \cos k_y- \notag \\
&&2t_{xx}^{11}(\cos 2k_x - \cos 2k_y) +
4t_{xyy}^{11} \cos 2k_x \cos k_y \notag \\
&&+ 4t_{xxy}^{11}\cos 2k_y \cos k_x+4t_{xxyy}^{11}\cos 2k_x\cos 2k_y\\
\xi_{33}&=&2t_{x}^{33}(\cos k_x + \cos k_y) + 4t_{xy}^{33} \cos k_x
\cos k_y +\notag \\
&&2t_{xx}^{33} (\cos 2k_x + \cos 2k_y) \\
\xi_{44}&=&2t_{x}^{44}(\cos k_x + \cos k_y) +4t_{xy}^{44} \cos k_x
\cos k_y +\notag \\
&&2t_{xx}^{44}(\cos 2k_x + \cos 2k_y)\notag \\
&&+4t_{xxy}^{44}(\cos 2k_x \cos k_y + \cos 2k_y \cos k_x)+\notag \\
&&4t_{xxyy}^{44} \cos 2k_x \cos2k_y\\
\xi_{55}&=&2t_{x}^{55}(\cos k_x + \cos k_y) +2t_{xx}^{55}(\cos 2k_x
+ \cos 2k_y)+\notag \\
&&4t_{xxy}^{55}(\cos 2k_x \cos k_y +\cos 2k_y \cos k_x)\notag \\
&&+4t_{xxyy}^{55}\cos 2k_x \cos 2k_y\\
\xi_{12}&=&4t_{xy}^{12}\sin k_x \sin k_y +\notag \\
&&4t_{xxy}^{12}(\sin 2k_x
\sin k_y +\sin 2k_y \sin k_x)+\notag \\
&&4t_{xxyy}^{12}\sin 2k_x \sin 2k_y\\
\xi_{13}&=&2it_{x}^{13}\sin k_y+4it_{xy}^{13}\sin k_y \cos k_x -\notag \\
&&4it_{xxy}^{13}(\sin 2k_y \cos k_x - \cos 2k_x \sin k_y)\\
\xi_{23}&=&2it_{x}^{13}\sin k_x+4it_{xy}^{13}\sin k_x \cos k_y -\notag \\
&&4it_{xxy}^{13}(\sin 2k_x \cos k_y - \cos 2k_y \sin k_x)\\
\xi_{14}&=&2it_{x}^{14} \sin k_x + 4it_{xy}^{14} \cos k_y \sin k_x +\notag \\
&&4it_{xxy}^{14} \sin 2k_x \cos k_y\\
\xi_{24}&=&-2it_{x}^{14} \sin k_y + 4it_{xy}^{14} \cos k_x \sin k_y
-\notag \\
&&4it_{xxy}^{14} \sin 2k_y \cos k_x\\
\xi_{15}&=&2it_{x}^{15}\sin k_y -4it_{xy}^{15}\sin k_y \cos k_x\notag \\
&&-4it_{xxyy}^{15} \sin 2k_y \cos 2k_x\\
\xi_{25}&=&-2it_{x}^{15}\sin k_x +4it_{xy}^{15}\sin k_x \cos k_y\notag \\
&&+4it_{xxyy}^{15} \sin 2k_x \cos 2k_y\\
\xi_{34}&=&4t_{xxy}^{34}(\sin 2k_y \sin k_x - \sin 2k_x \sin k_y)\\
\xi_{35}&=&2t_{x}^{35}(\cos k_x - \cos k_y)+\notag \\
&&4t_{xxy}^{35}(\cos 2k_x
\cos k_y -\cos 2k_y \cos k_x)\\
\xi_{45}&=&4t_{xy}^{45} \sin k_x \sin k_y + 4t_{xxyy}^{45} \sin 2k_x
\sin 2k_y
\end{eqnarray}
\par
 
The values of the hopping parameters $t_{\alpha}^{ij}$ are explicitly 
provided in Ref.~\onlinecite{graser}.


\begin{thebibliography}{10}
\bibitem{com} The value of $V$ has been chosen so that the gaps can be seen in
a scale where the whole bandwidth range is displayed.
\bibitem{foot} The five-orbital model being considered here 
has strong hybridization 
among the five orbitals and, as a result, the change of basis matrix $U$ does not
have a block diagonal 
structure for all values of the momentum. If this were the case, 
it would be possible to obtain a pure intraband pairing arising from intraorbital
pairing with orbital dependent strength. Orbitals that do not hybridize with 
each other could have different pairing strengths, which in turn may produce 
FSs with independent gapas it is the case in MgB$_2$\cite{mg} and
NdSe$_2$\cite{nd}.
\bibitem{foot3} Notice that linear combinations of 
form factors that create pairs of different sizes would be allowed by
symmetry considerations. However, since the spatial extension of the pairs
can be determined from experiments, we consider in this work pairing operators
that create pairs of a fixed size as given in Eq.~\ref{3}.
\bibitem{foot2} This actually results from the symmetry of the pnictides since 
among the four structural families 
constituting the iron-based superconducting pnictides the 11, 111, and 1111 are
characterized by a nonsymmorphic space group.\cite{eschrig} In this kind of 
systems all energy bands corresponding to elementary band representations 
are necessarily composite and connected.\cite{zac}
\end{thebibliography}
\end{document}